\documentstyle[prl,epsf,aps]{revtex}

 \newcommand \be {\begin{equation}}
\newcommand \bea {\begin{eqnarray} \nonumber }
\newcommand \ee {\end{equation}}
\newcommand \eea {\end{eqnarray}}
 
 \newcommand \bi {\bibitem}
\newcommand \s {\sigma}

\input{epsf}
\begin{document}
\twocolumn[\hsize\textwidth\columnwidth\hsize\csname@twocolumnfalse\endcsname
\title{Solving the Schr\"oedinger equation for the Sherrington-Kirkpatrick
model in a transverse field} \author{David Lancaster$^{(*)}$ 
and Felix Ritort$^{(**)}$}
\address{(*) Van der Waals-Zeeman Laboratorium\\
(**) Institute of Theoretical Physics\\ University of Amsterdam\\
Valckenierstraat 65\\ 1018 XE Amsterdam (The Netherlands).\\ E-Mail:
lancaste@phys.uva.nl,ritort@phys.uva.nl}

\date{\today}
\maketitle

\begin{abstract}
By numerically solving the Schr\"oedinger equation for
small sizes we investigate the quantum critical point
of the infinite-range Ising spin glass in a transverse field at
zero temperature. Despite its
simplicity the method yields accurate information on the value of the
critical field and critical exponents. We obtain $\Gamma_c=1.47\pm 0.01$
and check that exponents are in agreement with analytical approaches.
\end{abstract} 

\vfill

\vfill

\twocolumn
\vskip.5pc] 
\narrowtext

There has recently been renewed interest in the study of quantum phase
transitions in disordered systems\cite{QUANTUM_REVIEWS}. In particular,
Ising spin glass models in a transverse field are simple systems 
in which to study the effect of competition between randomness and quantum
fluctuations.  The case of infinite-range models is especially
interesting because they show non-trivial quantum phase transitions yet
are  to some extent amenable to analytical computations. The canonical
example in this family of models is the quantum Sherrington-Kirkpatrick
(SK) model in a transverse field. At zero transverse field this reduces to
the usual SK model which has a finite temperature
transition to low temperature phase where replica symmetry is broken
\cite{BOOKS}.  As the transverse field is turned on, spin-glass ordering
occurs at lower temperatures and above a certain critical field the
spin-glass order is completely suppressed at the expense of ordering in
the transverse direction.  Our understanding of this model was
significantly extended by the non-perturbative analysis of Miller and
Huse \cite{MIHU} and also by the different approach of Ye, Sachdev and
Read \cite{YESARE}.  The critical behavior is now well established, and
values for exponents, predictions for logarithmic corrections and
estimates of the value of the critical field are known.  The model is
therefore well adapted as a testing ground for numerical methods to
investigate quantum phase transitions. From this point of view, the
phase diagram of the quantum SK model in a transverse field, and its
zero temperature critical behavior have been studied using numerical
techniques such as spin summation \cite{GOLA}, perturbation expansions
\cite{ISYA} and quantum Monte Carlo methods \cite{RA}.  It is the
purpose of this letter to introduce a new numerical approach based on
the intuitive method of directly solving the Schr\"oedinger equation for
finite systems.  Despite its simplicity, this method is able to give
quantitative information on the value of the critical field and critical
exponents even for the very small size systems we consider.

The SK model in a transverse field is defined by the Hamiltonian,
\be {\cal H}={\cal H}_0-\Gamma {\cal M}_x=
-\sum_{i<j}J_{ij}\s_i^z\s_j^z-\Gamma\sum_i\s_i^x
\label{eq1}
\ee
where $\s_i^z,\s_i^x$ are Pauli spin matrices and $\Gamma$ is the
the transverse field.  The indices $i,j$ run from 1 to $N$ where $N$ is
the number of sites. The $J_{ij}$ are  Gaussian
distributed random variables with zero mean and $1/N$ variance. 
${\cal H}_0$ is the term 
we call the interaction energy, while ${\cal M}_x$ stands for the
magnetisation in the transverse direction.

We propose to study this model by the direct method of numerically
solving the real time Schr\"oedinger dynamics,
\be
i\frac{\partial \vert\psi\rangle}{\partial t}= {\cal H} \vert\psi\rangle
\label{eq2}
\ee
The wave function, $\vert\psi(t)\rangle$, of the system at time $t$ can
be written as a linear combination of basis states,
\be
\vert\psi(t)\rangle=\sum_{\nu=1}^{2^N} a_{\nu}(t) |\nu\rangle
\label{eq3}
\ee We have chosen the basis states, $\lbrace |\nu\rangle;
\nu=1,..,2^N\rbrace$ to be eigenstates of each
of the spin operators $\lbrace\s_z^i;i=1,N\rbrace$, 
$|\nu\rangle = |s_1,s_2\dots s_N\rangle$. This choice gives a
geometric meaning to equation (\ref{eq2}) because these eigenstates can
also be interpreted as the vertices of a unit hypercubic cell of
dimension $N$.  Each vertex of this hypercubic Hilbert space is assigned
a label $\nu$ and a corresponding complex variable $a_{\nu}$, which
together define the state of the system. This geometric picture can
also be used to understand the action of the Hamiltonian operator on the
basis states $|\nu\rangle$.
The action of the first term in equation (\ref{eq1}) on the state
$|\nu\rangle$ is diagonal with
eigenvalue $E^0_\nu$ which is precisely the energy of the classical
SK model in that state ($E^0_\nu=\langle \nu\vert{\cal
H}_0\vert\nu\rangle$).  The operator $\s_x^i$ acting on a given
eigenstate $|\nu\rangle$ changes the value of one spin which corresponds to an
adjacent vertex of the hypercube. The dynamical equations for the
$a_{\nu}$ become, \be i \frac{\partial a_{\nu}(t)}{\partial t}=
E^0_{\nu}a_{\nu}(t)-\Gamma\sum_{\mu(\nu)} a_{\mu}(t)
\label{eq4}
\ee
where $\mu(\nu)$ are nearest neighbours to the vertex $\nu$ in 
the hypercubic cell. The geometric picture also facilitates
efficient computer code for this problem.

We wish to calculate thermodynamic properties of the Hamiltonian
(\ref{eq1}) at zero temperature.
Such information
could be obtained by finding the static ground state of the Hamiltonian
${\cal H}$. However, because we want this ground state for a range of 
transverse fields, it is convenient to use a
dynamical procedure. At large  $\Gamma$ the Hamiltonian
can be diagonalised and the ground state is given by:
$a_{\nu}(t=0)=1/2^{\frac{N}{2}}$. Starting from this 
configuration we reduce the transverse field adiabatically
slowly, thus ensuring that the system remains in its ground
state. This procedure is recommended by its simple
physical interpretation but is not the most efficient
method that could be envisioned. For example the phase of the
wavefunction is not of interest, and for large systems a
gain in speed could be achieved by some less direct method.

We have numerically integrated eq.(\ref{eq4}) for different values of $N$
ranging from $N=2$ to $N=13$, using a simple Euler algorithm.
The value of the time step can be fixed by testing the conservation
of energy for some excited state at fixed $\Gamma$. For the ground 
state we can be less careful and we choose a time step $dt=0.01$.
The transverse field is allowed to decrease linearly 
from $\Gamma=3$ down to $\Gamma=0$. We find that a total time
of 100 units (amounting to 10000 integration steps) gives sufficiently
slow variation of $\Gamma$ for the adiabatic theorem to hold.
The method has also been checked against the analytic solution
of the model for $N=2$.
The errors from the discretisation of the Schr\"oedinger equation,
from the adiabatic approximation and from the finite initial value of 
$\Gamma$ are therefore well under control.
The main source of error 
comes from sample to sample fluctuations. Data was averaged
over many samples; ranging from 50000 for the smallest systems ($N=3,4,5$)
to 3000 for the largest sizes ($N=10,11,13$). We have also considered
100 samples at $N=17$ to confirm the tendency of the data, but have not
used these points in our fits due to the large errors.

\begin{figure}
\begin{center}
\leavevmode
\epsfysize=230pt{\epsffile{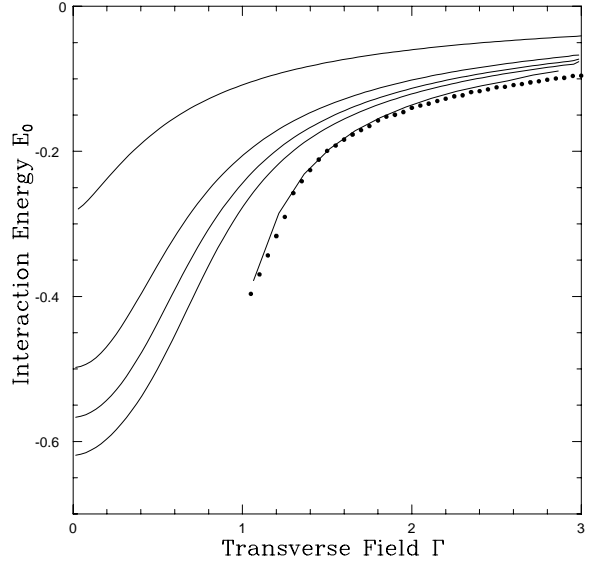}}
\end{center}
  \protect\caption[1]{ Interaction energy against tranverse field for
(from top to bottom) $N=2$ (analytic curve), $N=5$ (50000 samples),
$N=8$ (30000 samples) and $N=13$ (3000 samples). Errors are not
shown, but are always less than $10^{-3}$.
The lowest continuous
line is the extrapolated data, $E_0(\infty)$, and the points have been
obtained by quantum Monte Carlo methods \cite{RA}.   
\protect\label{FIG1} }
\end{figure}
\begin{figure}
\begin{center}
\leavevmode
\epsfysize=230pt{\epsffile{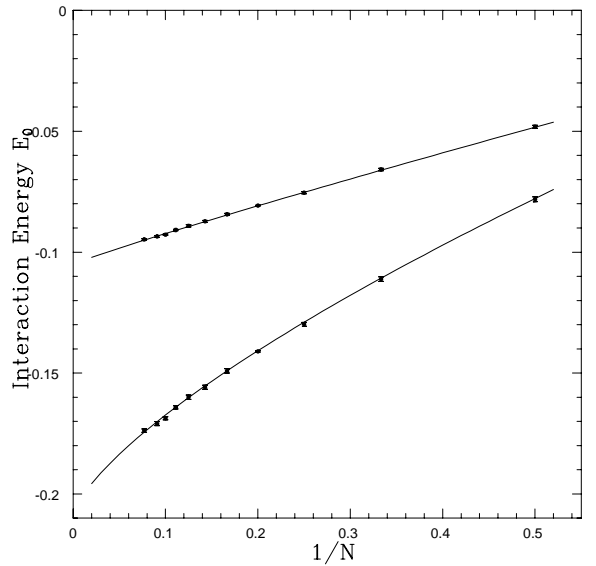}}
\end{center}
  \protect\caption[1]{
Interaction Energy against $1/N$. At $\Gamma = 2.5$ in the
QP phase (top) and $\Gamma = 1.45$ near the QC point (bottom). 
The points have error bars and the continuous curves show the
power law fits.
    \protect\label{FIG2}
  }
\end{figure}

The simplest variable is the interaction energy, $E_0=\langle 0|{\cal
H}_0|0\rangle=\sum_{\nu}E_{\nu}a_{\nu}a_{\nu}^*$.  It is plotted in
figure 1 as a function of $\Gamma$ for several different sizes. In
figure 2 we show $E_0$ for three different values of $\Gamma$ as a
function of $1/N$. The different values of $\Gamma$ are $\Gamma=2.5$ in
the quantum paramagnetic phase (QP) and $\Gamma=1.45$ near the quantum
critical (QC) point (see below). Data has been fitted with the least
squares method to a power law of the type $E_0(N)=E_0(\infty)+a N^{-b}$
\cite{FSS,YO}.  In the $QP$ phase $b\simeq 0.93$, so corrections are
essentially $1/N$ as expected. Close to the QC point we find
$b=0.73\pm0.02$ in agreement with the expected mean-field exponent $b=1
+ \frac{z}{d_u} +\frac{1}{\nu d_u}=3/4$ where $\nu$ is the correlation
length exponent ($\nu = 1/4$), $z$ is the dynamical exponent ($z=2$) and
$d_u$ is the upper critical dimension ($d_u =8$). It
is remarkable that even very small size systems fit on the curve.  This
data is summarized in figure 1 where we have also shown the best fit 
parameters $E_0(\infty)$ as a function of $\Gamma$ in the region 
$\Gamma > 1.2$.  The points are numerical data
obtained by independent Quantum Monte Carlo calculations \cite{RA} and
show reasonable agreement with the extrapolated values.
At smaller $\Gamma$, $b$ decreases and it is no longer possible to 
ignore sub-leading corrections; nonetheless, at $\Gamma = 0$ we find
$E_0(\infty) \simeq -0.763$ in agreement with the theory\cite{BOOKS}.

\bigskip

These results give us confidence in the method and encourage us
to investigate the transition more closely.
Because of the spin glass nature of the transition, a clear
signal does not appear in the more ordinary thermodynamic 
functions. A divergence will occur in the non-linear susceptibility
and we have also seen
a peak in a Binder-like parameter for the kurtosis of the
sample to sample distribution of the interaction energy.
These observations suggest a transition in the region of
$\Gamma \sim 1.5$ as expected, but are not the best way to 
obtain accurate information.
To determine the critical field and critical exponents
we consider the longitudinal susceptibility associated to the
magnetisation ${\cal M}_z=\sum_i\s_i^z$. 
It can be shown \cite{BM} that the longitudinal susceptibility ($\chi$)
for the SK model is precisely equal to 1 at the QP-QG boundary.
To this end we have numerically computed $\chi$. 
The usual quantum mechanical formula,
\be
\chi = \sum_{n\ne 0}\frac{|\langle n|{\cal M}_z|0\rangle |^2}{E_n-E_0},
\label{eq5}
\ee
where $|n\rangle$ denotes the energy eigenstate with energy $E_n$,
is not adapted to our needs because it requires knowledge of the
full spectrum. Instead, we solve the problem directly 
by applying a longitudinal magnetic field $h$ small enough to be
in the linear response regime. The susceptibility is computed as
$\chi = {\langle 0'|{\cal M}_z|0'\rangle}/{h}$ 
where $|0'\rangle$ stands for the ground
state in the presence of the field $h$. 
We need only solve the Schr\"oedinger equation once for the
perturbed Hamiltonian ${\cal H} + h {\cal M}_z$, since
the magnetisation of the 
unperturbed problem (without magnetic field) is strictly
zero, as follows from a quantum mechanical symmetry. For zero transverse
field this symmetry is the spin reversal symmetry of the classical
SK model. In the thermodynamic limit $N \rightarrow \infty$
only one state is selected, but for our finite systems
the wavefunction always contains a mixture of opposite
magnetisation states. In the perturbed case this symmetry is lost 
and at small transverse fields the action of the perturbation is to
shift the wavefunction to a single magnetised state. This 
quantum tunneling introduces a new time scale into the problem and
the parameters we use in solving the Schr\"oedinger equation
should be re-examined. 
We have checked that the parameters we use give small
errors (less than the errors from sample fluctuations) around
the region of criticality.

\begin{figure}
\begin{center}
\leavevmode
\epsfysize=230pt{\epsffile{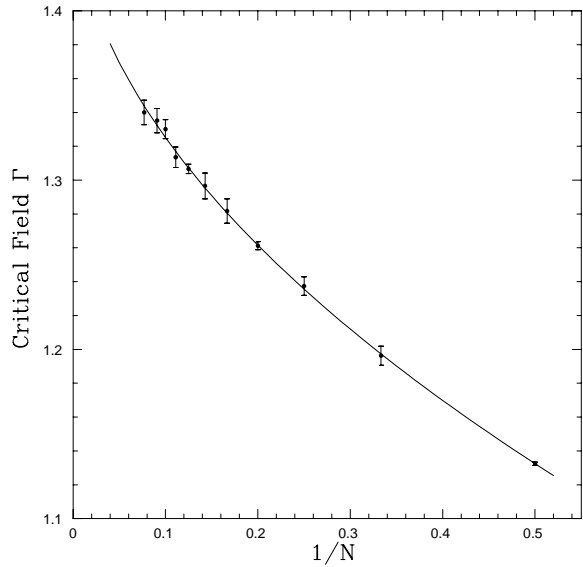}}
\end{center}
  \protect\caption[1]{
Critical Transverse field against $1/N$. 
    \protect\label{FIG3}
  }
\end{figure}

The actual value of the perturbing field $h$ can be taken in a wide
range without affecting results, and we present data for $h$ as small
as $10^{-7}$.  Using the exact condition for criticality we define the
critical transverse field by $\chi(\Gamma) =1$, and in figure 3 we
plot this $\Gamma$ against $1/N$.  Fitting the data to a power law
behavior of the type $\Gamma=\Gamma_c+a N^{-b}$ we find
$\Gamma_c=1.47\pm0.02$, $a=-0.485\pm 0.002, b=0.53\pm 0.02$ in good
agreement with the results obtained in perturbation theory by Ishii
and Yamamoto \cite{ISYA} ($\Gamma_c=1.506$) and with the result
obtained by Miller and Huse \cite{MIHU} ($\Gamma_c=1.46\pm0.01$).  The
coefficient $b$ is very close to the expected value $b=\frac{1}{\nu
d_u}=1/2$. If we consider
the scaling at $\Gamma= 1.47$ we find that $\chi$ converges to 1
as $N^{-c}$ with a value of the exponent $c\simeq 0.29$
compatible with $c=\frac{2}{d_u}=1/4$.  
The data collapse in the scaling region is
shown in figure 4 where $(1-\chi)N^{\frac{2}{d_u}}$ is plotted as a
function of $N^{\frac{1}{\nu d_u}}(\Gamma-\Gamma_c)$. The collapse is
good and confirms the expected values $\nu=1/4$ and $d_u=8$. Even the
result for $N=2$ lies close to the collapse line.  Considering the
simplicity of the method and the small sizes considered the matching
with data reported in the literature is impressive.  This is
particularly the case since logarithmic corrections are known to be
present \cite{MIHU,YESARE}.

\begin{figure}
\begin{center}
\leavevmode
\epsfysize=230pt{\epsffile{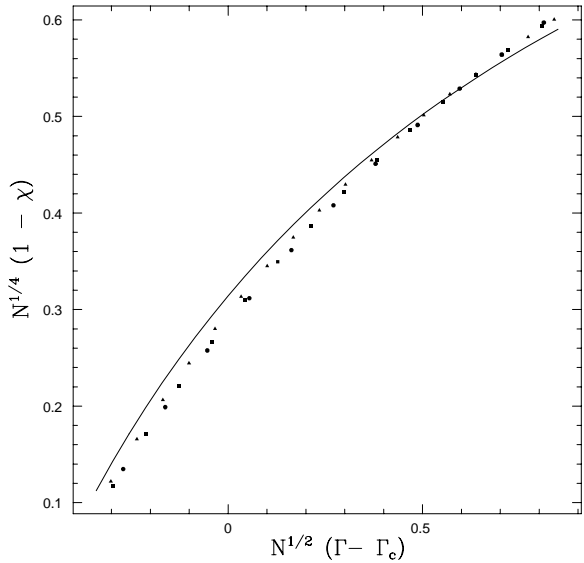}}
\end{center}
  \protect\caption[1]{
Data collapse for $N=5$ (triangles), $N=8$ (squares) and
$N=13$ (circles). Shown for the range $1.2<\Gamma <2.0$ using our value 
of $\Gamma_c=1.47$.
The continuous line is the analytical result for $N=2$.
    \protect\label{FIG4}
  }
\end{figure}

In summary, a new and simple numerical method yielding good estimates of
the critical field and confirming the critical exponents for
the quantum phase
transition of the SK model has been reported. The method consists in
solving the Schr\"oedinger equation for small sizes and computing
expectation values in the ground state of the system.  Quite remarkably,
the system is within the scaling region even for very small sizes.  The
application of this method to other disordered systems such
as the random orthogonal model \cite{RIT} and the quantum Potts glass
\cite{SE} would be very welcome.

{\bf Acknowledgments.} 
F.R is grateful to FOM for financial support.  We acknowledge
conversations on this and related subjects with Th.~M.~Nieuwenhuizen.

\end{document}